\documentclass{achemso}
\usepackage[T1]{fontenc}
\usepackage{graphicx}
\usepackage{babel}
\usepackage{hyperref}
\makeatletter
\usepackage{gensymb}
\usepackage{geometry}
\usepackage{array}
\usepackage{enumitem}
\usepackage{diagbox}
\usepackage{amsmath}
\usepackage{multirow} 
\usepackage[table,xcdraw]{xcolor}
\usepackage{tikz}
\usepackage{threeparttable} 

\title{AI-assisted design of chemically recyclable polymers for food packaging}
\author{Brandon K. Phan}
\affiliation{School of Materials Science and Engineering, Georgia Institute of Technology, Atlanta, Georgia 30332, United States}

\author{Chiho Kim}
\affiliation{School of Materials Science and Engineering, Georgia Institute of Technology, Atlanta, Georgia 30332, United States}

\author{Janhavi Nistane}
\affiliation{School of Materials Science and Engineering, Georgia Institute of Technology, Atlanta, Georgia 30332, United States}

\author{Wei Xiong}
\affiliation{School of Materials Science and Engineering, Georgia Institute of Technology, Atlanta, Georgia 30332, United States}

\author{Haoyu Chen}
\affiliation{School of Chemical and Biomolecular Engineering, Georgia Institute of Technology, Atlanta, Georgia 30332, United States}

\author{Woo Jin Jang}
\affiliation{School of Chemical and Biomolecular Engineering, Georgia Institute of Technology, Atlanta, Georgia 30332, United States}

\author{Farzad Gholami}
\affiliation{School of Materials Science and Engineering, Georgia Institute of Technology, Atlanta, Georgia 30332, United States}

\author{Yongliang Su}
\affiliation{School of Chemical and Biomolecular Engineering, Georgia Institute of Technology, Atlanta, Georgia 30332, United States}

\author{Jerry Qi}
\affiliation{School of Materials Science and Engineering, Georgia Institute of Technology, Atlanta, Georgia 30332, United States}

\author{Ryan Lively}
\affiliation{School of Chemical and Biomolecular Engineering, Georgia Institute of Technology, Atlanta, Georgia 30332, United States}

\author{Will Gutekunst}
\affiliation{School of Chemical and Biomolecular Engineering, Georgia Institute of Technology, Atlanta, Georgia 30332, United States}

\author{Rampi Ramprasad}
\affiliation{School of Materials Science and Engineering, Georgia Institute of Technology, Atlanta, Georgia 30332, United States}

\makeatletter
\email{rampi.ramprasad@mse.gatech.edu}
\makeatother
\begin{document}

\maketitle

\begin{abstract}

Polymer packaging plays a crucial role in food preservation but poses major challenges in recycling and environmental persistence. To address the need for sustainable, high-performance alternatives, we employed a polymer informatics workflow to identify single- and multi-layer drop-in replacements for polymer-based packaging materials. Machine learning (ML) models, trained on carefully curated polymer datasets, predicted eight key properties across a library of approximately 7.4 million ring-opening polymerization (ROP) polymers generated by virtual forward synthesis (VFS). Candidates were prioritized by the enthalpy of polymerization, a critical metric for chemical recyclability. This screening yielded thousands of promising candidates, demonstrating the feasibility of replacing diverse packaging architectures. We then experimentally validated poly($p$-dioxanone) (poly-PDO), an existing ROP polymer whose barrier performance had not been previously reported. Validation showed that poly-PDO exhibits strong water barrier performance, mechanical and thermal properties consistent with predictions, and excellent chemical recyclability ($\sim$95\% monomer recovery), thereby meeting the design targets and underscoring its potential for sustainable packaging. These findings highlight the power of informatics-driven approaches to accelerate the discovery of sustainable polymers by uncovering opportunities in both existing and novel chemistries.

\end{abstract}

\section{Introduction}

Food preservation is a contemporary challenge that requires materials providing both protection and sustainability. Appropriately optimized polymeric materials may serve as effective alternatives to conventional solutions. In particular, plastic packaging has fundamentally transformed global food supply chains by offering superior protective performance. These materials not only prolong product shelf life but also minimize food waste and support the worldwide distribution of perishable goods. Consequently, polymers have become widespread in modern industrial and consumer applications, forming the foundation of the packaging landscape.\cite{ebnesajjad2012plastic}

\begin{figure}[t!]
    \centering
    \includegraphics[width=1\columnwidth]{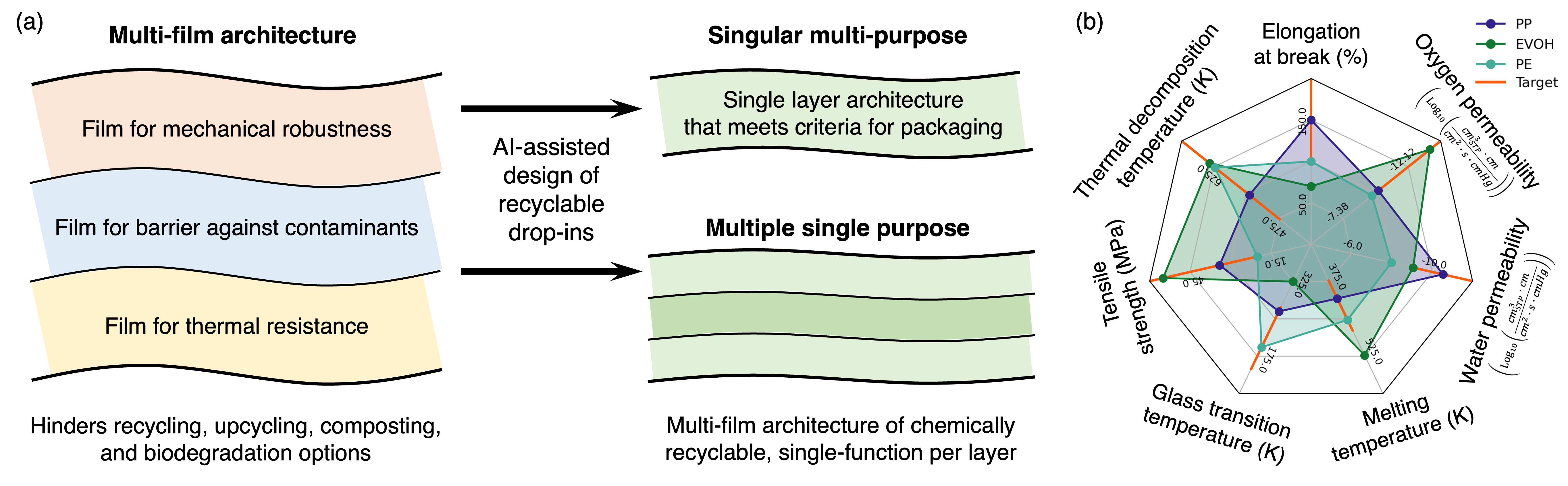}
\caption{Sustainable food packaging design strategy and performance targets. (a) Conventional food packaging with complex multi-layer designs where each layer performs a specific function, complicating chemical recycling, can be replaced with two simplified, chemically recyclable alternatives, including a single-layer multi-purpose polymer and a multi-layer structure where each layer is a single-function, chemically recyclable polymer. (b) The design criteria are compared against the property profiles of common polymers (PP, EVOH, PE), highlighting the performance gaps that the newly designed polymer candidates must fill.}
\label{fig:scheme}
\end{figure}

Despite these functional strengths, the very architecture and chemical persistence that make current packaging materials effective also represent major environmental challenges. Present-day packaging plastics often rely on a multi-layer architecture (Fig. \ref{fig:scheme}a), combining distinct, often non-compatible, layers for barrier, mechanical, and thermal functions. While effective, these materials are notorious for their persistence in landfills, fragmenting into microplastics that contribute to long-term environmental pollution \cite{hale2020global,andrady2011microplastics,andrady2017plastic}. Furthermore, this multi-layer composition poses a significant obstacle to recycling since the chemically distinct layers must be separated, a time- and resource-intensive process. The difficulty in finding suitable replacement materials that meet all necessary performance standards is highlighted by a comparison of the properties of currently used polymers, such as polypropylene (PP), ethylene vinyl alcohol copolymer (EVOH), and polyethylene (PE) (Fig. \ref{fig:scheme}b).

Recognizing these limitations, existing academic and industrial efforts aim to develop alternative packaging materials that address sustainable end-of-life scenarios while maintaining the necessary performance standards. Current sustainable strategies include replacing conventional plastics with those that can undergo efficient chemical recycling (through triggered reversible conversion between monomer and polymer) or complete biodegradation. Another more onerous design approach seeks the consolidation of the multi-film configuration into a single multi-functional layer, often referred to as a "drop-in replacement." This strategy requires optimizing barrier performance, mechanical strength, and thermal robustness simultaneously, a complex design challenge that renewable polymers often struggle to meet \cite{wu2021challenges}.

To accelerate the discovery of these high-performance, sustainable alternatives, this work leverages recent advances in materials informatics methods to accelerate the design and property prediction of new polymers \cite{audus2017polymer, batra2021emerging, chen2021polymer,tran2023informatics2, zhu2020polymer, barnett2020designing,wu2022rational,chen2020frequency}. We developed a comprehensive polymer informatics workflow that incorporates computational modeling and machine learning (ML) tools to identify candidates that address the specific performance shortcomings of sustainable polymers. Through this workflow, we identified potential chemically recyclable polymers that may be created via ring-opening polymerization (ROP) as potential drop-in replacements for both single multi-functional layers and current multi-layer architectures. This class of polymers with the ability to depolymerize and repolymerize not only reduces the need for new raw materials but also minimizes waste, thereby lessening the dependency on virgin resources and lowering the ecological footprint of polymer production. From this pool, we successfully synthesized and validated the key properties of poly($p$-dioxanone) (poly-PDO), a promising single-layer candidate \cite{saska2021polydioxanone,mathews2025polydioxanone,goonoo2015polydioxanone}. 

\section{Workflow}

\begin{figure}[t!]
    \centering
    \includegraphics[width=1\columnwidth]{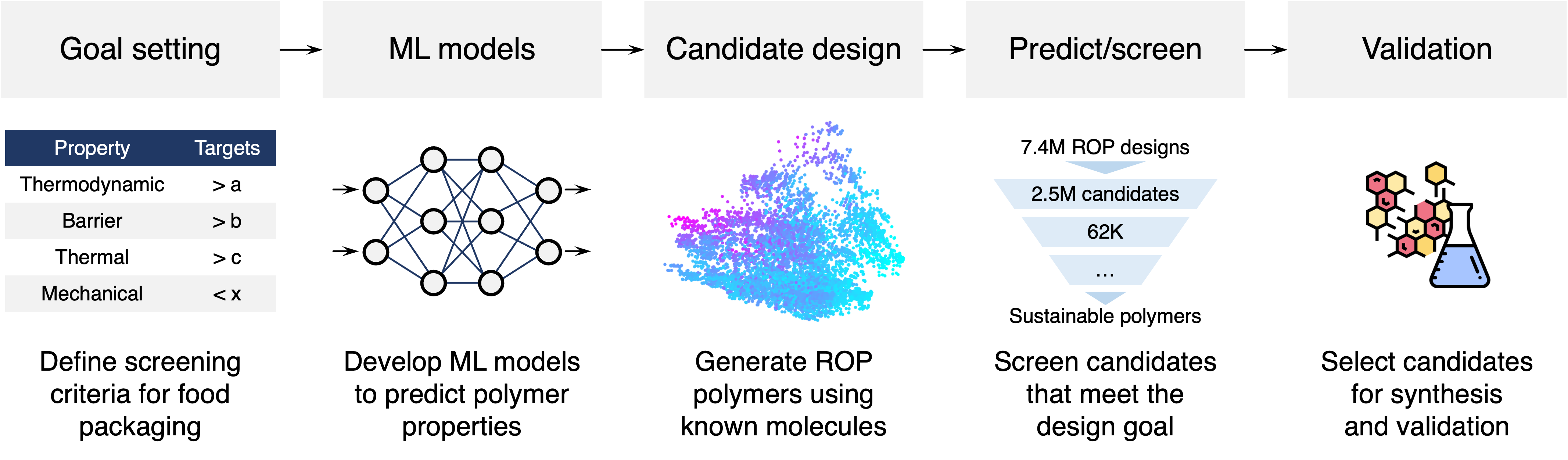}
\caption{AI-assisted design workflow for recyclable polymer discovery. A schematic overview of the data-driven methodology, illustrating the key stages: goal setting, defining target properties, developing predictive ML models, candidate design using virtual forward synthesis (VFS), predicting/screening by applying models to filter candidates, and experimental verification of promising cases.}
\label{fig:workflow}

\end{figure}

Our overall strategy follows a materials informatics workflow designed to accelerate the discovery of sustainable polymers \cite{kern2025informatics,tran2025polymer,nistane2025polymer,aklujkar2025rationally,schertzer2025ai,atasi2024design}. This approach, illustrated in Fig. \ref{fig:workflow}, systematically moves from defining target properties and curating necessary data to virtually synthesizing, predicting, and finally validating polymer candidates. All computational steps in this workflow, except for initial data collection and final experimental validation, were performed using PolymRize\texttrademark\ \cite{Matmerize_Inc}, a standardized software for molecular and polymer informatics.

The workflow begins by defining the target performance criteria necessary for a single-layer, multi-functional packaging material, guided by the common packaging polymers shown in Fig. \ref{fig:scheme}b. These essential properties, including barrier, thermal, and mechanical requirements, are outlined in Table \ref{table:criteria}. Following the definition of these goals, the necessary data for the eight critical properties were compiled. The data were collected from two distinct sources: high-fidelity experimental measurements (literature, handbooks, and online databases) and lower-fidelity computational results obtained from molecular dynamics (MD), Monte Carlo (MC), and density functional theory (DFT) calculations, as described in several prior articles \cite{phan2024gas,toland2023accelerated}. 

\begin{table}[t!]
    \centering
    \begin{threeparttable} 

        \caption{\textbf{Screening criteria for sustainable packaging polymers.}}
        \label{table:criteria}
        \begin{tabular}{ll} 
            \hline
            \textbf{Property} & \textbf{Desired target} \\
            \hline
            Enthalpy of polymerization & $-10$ to $-20$ $\text{kJ/mol}$ \\
            \hline
            Water vapor permeability at 25 $\degree$C & $<10^{-9.3}$ $\text{cm}^{3}_{\text{STP}}\cdot \text{cm}/(\text{cm}^{2}\cdot \text{s} \cdot \text{cmHg})$$^{(a)}$ \\
            \hline
            Oxygen permeability at 25 $\degree$C & $<10^{-10.2}$ $\text{cm}^{3}_{\text{STP}}\cdot \text{cm}/(\text{cm}^{2}\cdot \text{s} \cdot \text{cmHg})$$^{(a)}$ \\
            \hline
            Glass transition temperature & $< 298$ $\text{K}$ \\
            \hline
            Melting temperature & $> 373$ $\text{K}$ \\
            \hline
            Degradation temperature & $> 473$ $\text{K}$ \\
            \hline
            Elongation at break & $> 10^{2.17}$ \% \\
            \hline
            Tensile strength & $> 20$ $\text{MPa}$ \\
            \hline
        \end{tabular}
        
        \begin{tablenotes} 
            \item[a] 1 $\text{cm}^{3}_{\text{STP}}\cdot \text{cm}/(\text{cm}^{2}\cdot \text{s} \cdot \text{cmHg})$ = $10^{10}$ $\text{Barrer}$ \\
        \end{tablenotes}
    \end{threeparttable} 

\end{table}

With the datasets in place, predictive ML models were developed using Gaussian process regression for the enthalpy of polymerization and deep learning for all other properties, as described in previous work \cite{toland2023accelerated,doan2020machine}. Multi-task (MT) learning was strategically employed to improve accuracy by information fusion from different sources or by jointly learning correlated properties. Detailed information regarding the size and composition of the training data, the specific training algorithm configurations, and the resulting model performance metrics and parity plots is available in the Supplementary Information, and in previously published works \cite{phan2024gas}.

To efficiently explore the vast chemical space for sustainable solutions, we generated a large library of hypothetical, yet synthetically feasible, polymer designs. This Virtual Forward Synthesis (VFS) process \cite{kern2025informatics, shivank2025RxnChainer} was executed using the RxnChainer (Reaction Chainer) tool within the PolymRize\texttrademark\ platform \cite{Matmerize_Inc}. Developing such extensive and chemically accurate datasets is a non-trivial process, requiring expertise in both reaction chemistry and informatics to encode precise rules. RxnChainer streamlines this complex workflow by systematically leveraging the established ROP reaction to combine about 30 million commercially available molecules. This targeted process successfully yielded a robust virtual library of approximately 7.4 million polymer candidates previously discussed \cite{kern2025informatics} and released \cite{kern_git}, which served as the comprehensive search space for the subsequent screening efforts in this work.

The ML models were then used to predict the eight target properties for the virtually synthesized polymer library, followed by screening against the criteria defined in Table \ref{table:criteria}. Promising candidates were subsequently selected for experimental validation. The detailed synthesis and characterization procedures for the validated polymers are described in the Supplementary Information.

\begin{figure}[t!]
    \centering
    \includegraphics[width=0.9\columnwidth]{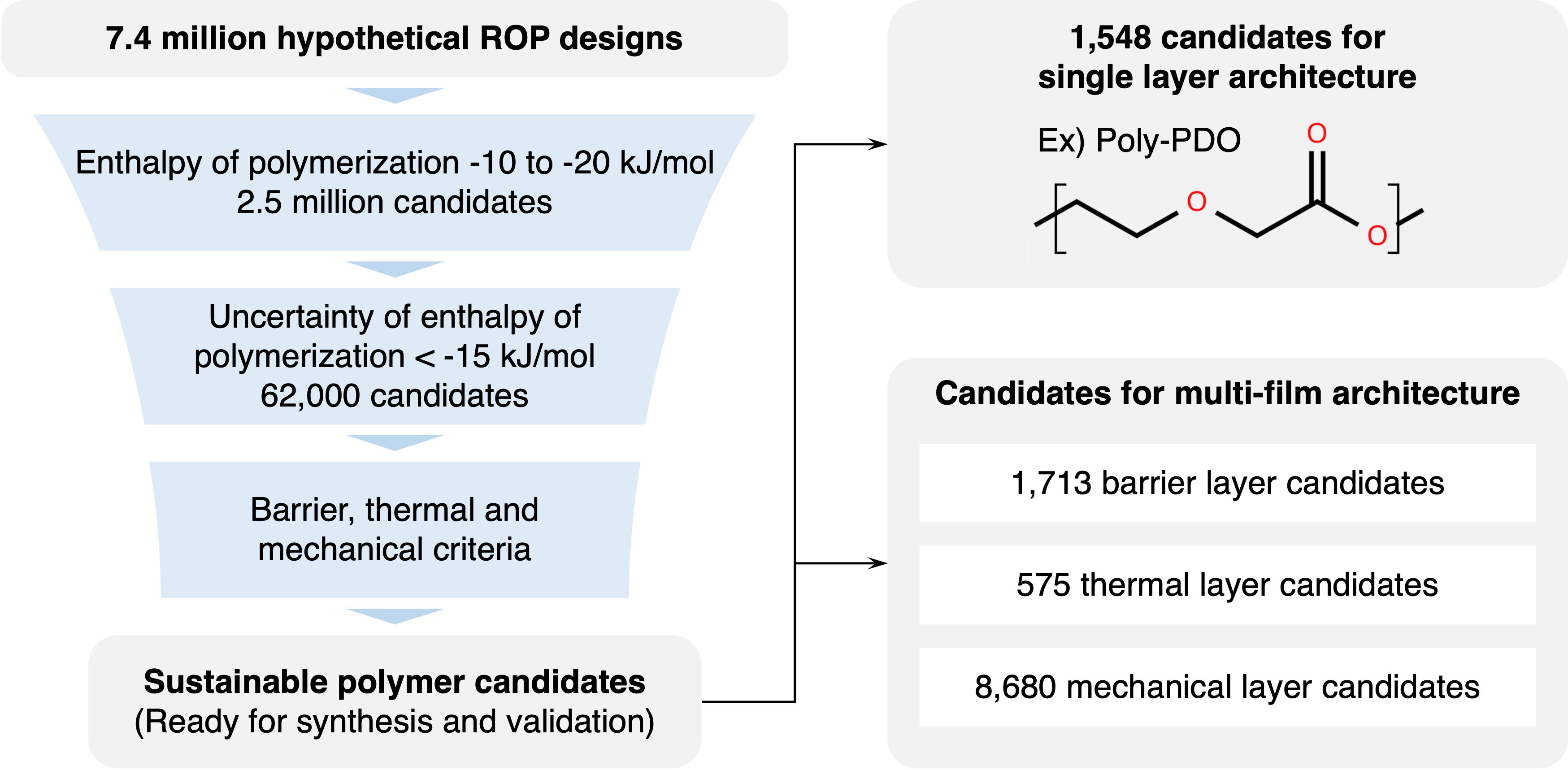}
\caption{Schematics for the multi-stage screening process to achieve sustainable polymer candidates in single- and multi-layer packaging architectures.}
\label{fig:screening}
\end{figure}

\section{Results and Discussion}

\begin{figure}[t!]
    \centering
    \includegraphics[width=1\columnwidth]{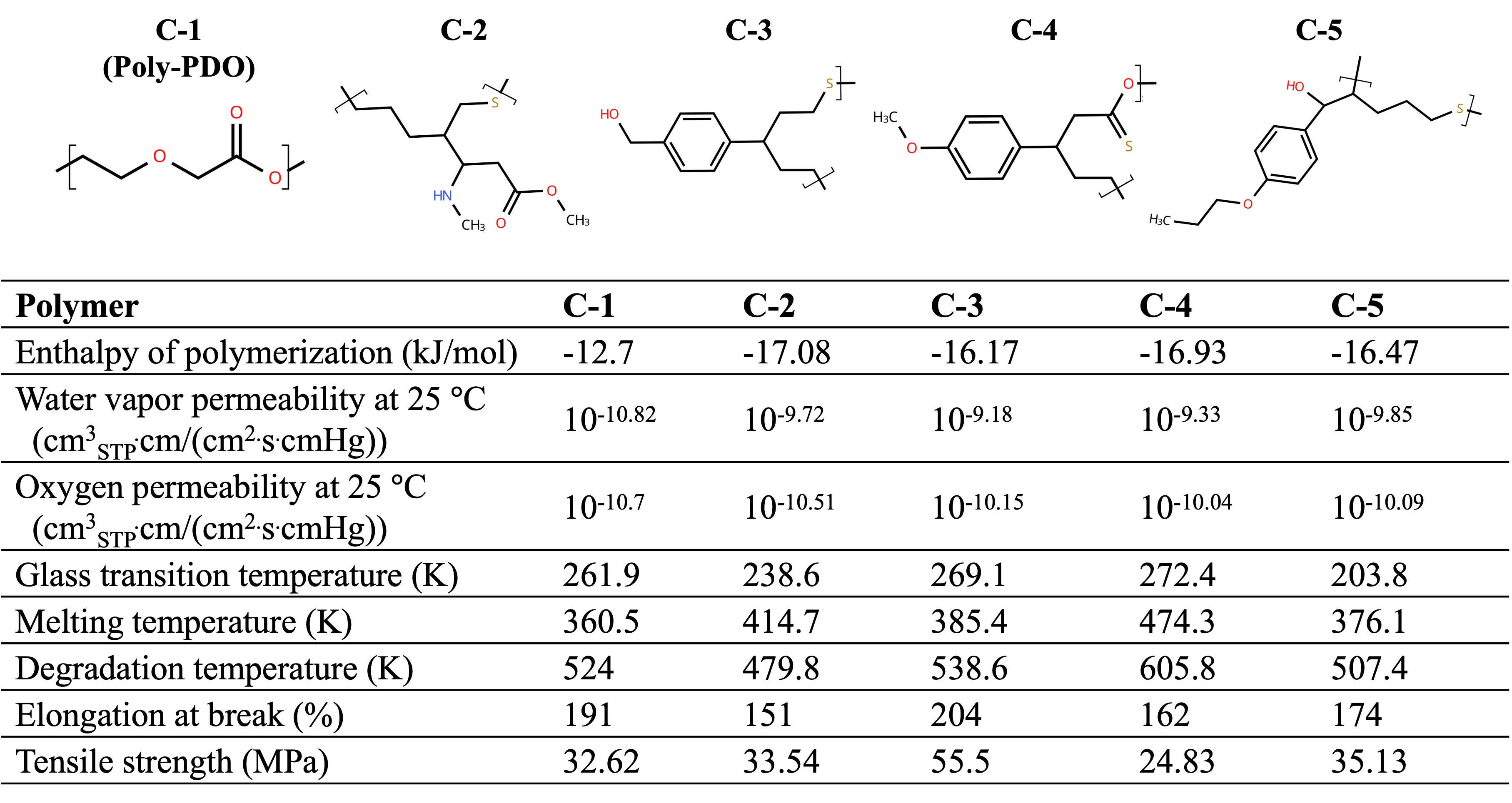}
\caption{Structure of monomers for polymerization, and predicted property values of five (of 1,548) candidates found as promising food packaging replacement polymers.}
\label{fig:candidates}
\end{figure}

Leveraging the predictive ML models, we rapidly predicted eight properties across the hypothetical ROP dataset (approximately 7.4 million candidates) and screened them against the criteria to identify single- and multi-layer drop-in replacements. The most critical filter was the enthalpy of polymerization, a key property for assessing chemical recyclability. From the initial 7.4 million population, about 62,000 candidates fell within the targeted enthalpy of polymerization range (-10 to -20 kJ/mol) and possessed low prediction uncertainty ($<$15 kJ/mol). The remaining thermal, mechanical, and barrier properties were then predicted for this filtered set. The systematic funneling process is illustrated in Fig. \ref{fig:screening}, and the results show the feasibility of finding replacements for both single and multi-layer architectures. The predicted properties and structures for a selection of promising single-layer replacement candidates are presented in Fig. \ref{fig:candidates}.

Based on its predicted properties and prior reports of successful synthesis in the literature, we designated poly($p$-dioxanone) (poly-PDO) as the lead candidate for experimental validation as a single-layer food packaging replacement. We synthesized poly-PDO through the ROP of $p$-dioxanone using Sn(Oct)$_2$ as a catalyst (Fig. \ref{fig:screening}), with the resulting structure confirmed by $^1$H nuclear magnetic resonance (NMR) spectroscopy (Fig. S2) and a molecular weight of 26.6 kg/mol. While poly-PDO's thermal and mechanical data were reported in the literature, essential gas and water vapor permeability values, which are critical for food packaging applications, have not been studied before. Our experimental validation focused on filling this critical data gap in the literature, along with performing comprehensive characterizations of all relevant physical properties.

The thermal properties were rigorously assessed using differential scanning calorimetry (DSC). The measurements confirmed that the glass transition temperature of 257 K and the melting temperature of 378 K are in close agreement with the ML predictions (261.9 K and 360.5 K, respectively) and established literature ranges. The instrumentation and setup parameters for the thermal characterization are detailed in the Supplementary Information (SI). This initial confirmation validated the predictive power of our ML models for key thermal traits. Table \ref{table:pdo} summarizes experimental results, directly comparing the ML predictions with our experimental measurements and available literature data across all eight target properties.

\begin{table}[t!]
    \centering
    \begin{threeparttable} 

        \caption{Comparison of poly-PDO predictions, measurements, and literature values. 
            \footnotesize \textsuperscript{a}\cite{libiszowski2004monomer}, \textsuperscript{b}\cite{lelkes2024comprehensive}, \textsuperscript{c}\cite{liu2021effect}, \textsuperscript{d}\cite{kang2022effects}, \textsuperscript{e}\cite{saska2021polydioxanone}
        }
        \label{table:pdo}
        \begin{tabular}{l c c c}
    \hline
    \textbf{Property} & \textbf{Prediction} & \multicolumn{2}{c}{\textbf{Measured}} \\
    \cline{3-4}
     &  & \textbf{This work} & \textbf{Literature} \\
    \hline
            Enthalpy of polymerization ($\text{kJ/mol}$)& $-12.7 \pm 3.3$ & - & $-13.8^{a}$ \\
            \hline
            Water vapor permeability at 25 $\degree$C  & $10^{-10.82 \pm 0.2}$ & $10^{-10.7}$ & - \\
            ($\text{cm}^{3}_{\text{STP}}\cdot \text{cm} / (\text{cm}^{2}\cdot \text{s} \cdot \text{cmHg}$)) &  &   &  \\
            \hline
            Oxygen permeability at 25 $\degree$C  & $10^{-10.7 \pm 0.24}$ & $10^{-9.0}$ & -  \\
            ($\text{cm}^{3}_{\text{STP}}\cdot \text{cm} / (\text{cm}^{2}\cdot \text{s} \cdot \text{cmHg}$)) &   &   &  \\
            \hline
            Glass transition temperature (K) & $261.9 \pm 38$ & $257$ & $261$ to $263^{b,c}$ \\
            \hline
            Melting temperature (K) & $360 \pm 29$ & $378$ & $363$ to $397^{b,c}$ \\
            \hline
            Degradation temperature (K) & $524 \pm 55$ & $487$ & - \\
            \hline
            Elongation at break (\%) & $10^{2.28 \pm 0.1}$ & $10^{0.4}$ & $10^{1}$ to $10^{2.77c,d}$ \\
            \hline
            Tensile strength ($\text{MPa}$) & $32.6 \pm 6$ & $3$ & $3.9$ to $60^{d,e}$ \\
            \hline
        \end{tabular}
    \end{threeparttable} 

\end{table}

Poly-PDO demonstrated excellent barrier performance for water vapor, meeting the target goal. The ML prediction and experimental measurement aligned closely for water vapor permeability, classifying poly-PDO as a high-grade barrier that comfortably surpasses the set threshold. However, the oxygen barrier property remains a limitation, exhibiting one order higher permeability than the targeted design goal. Despite this constraint, poly-PDO is still well-suited for applications like fruit and salad packaging, where stringent oxygen barrier requirements are less critical.\cite{wu2021challenges}

The agreement between the predicted and experimentally measured mechanical properties (elongation at break and tensile strength) showed a discrepancy of approximately one order of magnitude (Table \ref{table:pdo}). While the predicted values met the design targets, the experimental results fell short of these expectations. This disparity may be due to several factors, including the broad spread in reported mechanical properties across studies, which suggests that our own measurements may require further optimization of the polymer samples to obtain accurate values. Such variability also indicates that the dataset used to train the ML model might include inconsistent or noisy data, emphasizing the need to assess data quality and incorporate more reliable measurements to improve model performance. Nevertheless, the predicted values remain within the range of previously reported experimental results, indicating that the overall trends are captured reasonably well.

To confirm the chemical recyclability, we characterized the chemical depolymerization of poly-PDO using DBU in benzene solution at 79 $\degree$C. This process demonstrated remarkable efficiency, achieving over 95\% monomer recovery and approaching quantitative yields of $\sim$100\% within six hours. The nearly complete recovery of the PDO monomer was confirmed by NMR spectroscopy (Fig. S3). More details on the experimental procedures and additional data are provided in the Supporting Information. This robust chemical recyclability, coupled with the fact that poly-PDO is also recognized as a biodegradable material \cite{choi2020physical,nishida2000microbial,kang2022effects,jin2013biodegradation}, highlights its potential to offer multiple sustainable pathways for food packaging, encompassing both circular economy strategies and end-of-life biodegradation options.

\section{Conclusion}

We successfully implemented a polymer informatics workflow, leveraging predictive ML models and digital reactions to accelerate the search and discovery of sustainable packaging materials. This AI-assisted and data-driven approach systematically screened the initial dataset of approximately 7.4 million ROP polymers, down-selecting to a pool of promising single- and multi-layer polymer film architectures.

The approach was validated through the experimental synthesis and characterization of poly-PDO, a rediscovered existing ROP polymer. We confirmed its strong potential by measuring high-grade water barrier performance and verifying its excellent chemical recyclability (over 95\% monomer recovery in six hours). The strong agreement between our experimental measurements and ML predictions for poly-PDO affirms the robustness of our predictive models. The mechanical properties showed discrepancies between predicted and measured values, highlighting the need for further optimization of polymer samples and refinement of the ML models. Addressing such shortcomings will be critical to extend both experimental optimization and predictive reliability, in a quest to design truly useful and practical polymers for a sustainable world.

\section{Acknowledgments}
This work is financially supported by the Office of Naval Research through a multidisciplinary university research initiative (MURI) grant N00014-20-1-2586 and the National Science Foundation through a NSF-SPEED grant 2515411. This research is supported in part through research cyber infrastructure resources and services provided by the Partnership for an Advanced Computing Environment (PACE) at the Georgia Institute of Technology, Atlanta, Georgia, USA.\cite{PACE} 

\newpage
\bibliography{refs}
\end{document}


\maketitle

\newpage
 
\section{Screening criteria}

\begin{table}[h!]
    \centering
    \includegraphics[width=0.8\columnwidth]{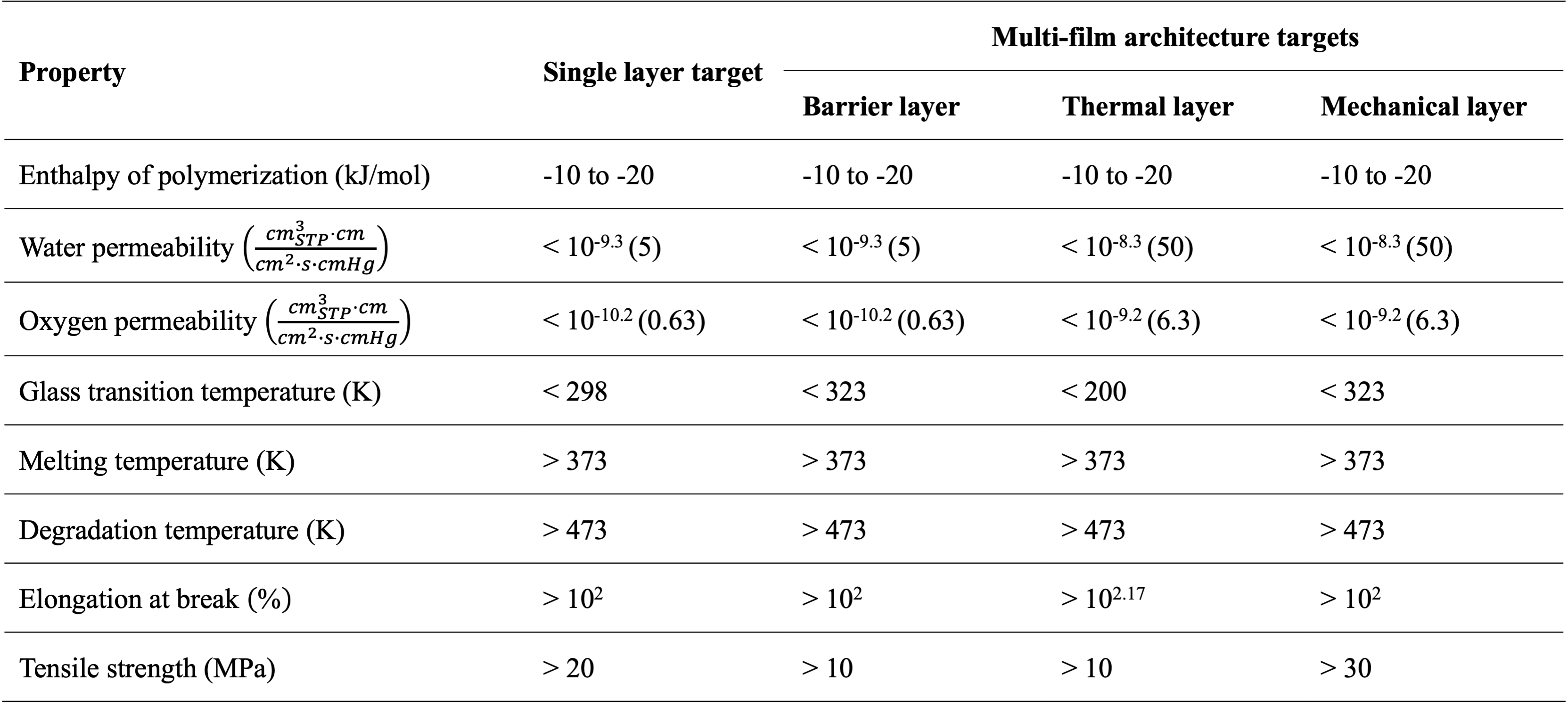}
    \caption{Property requirements set for single, barrier, thermal, and mechanical layer replacements. Permeability values enclosed in parenthesis are in units of Barrer.}\label{fig:reqs2}
\end{table}

\section{Machine learning models}

\begin{table}[h!]
    \centering
    \includegraphics[width=0.8\columnwidth]{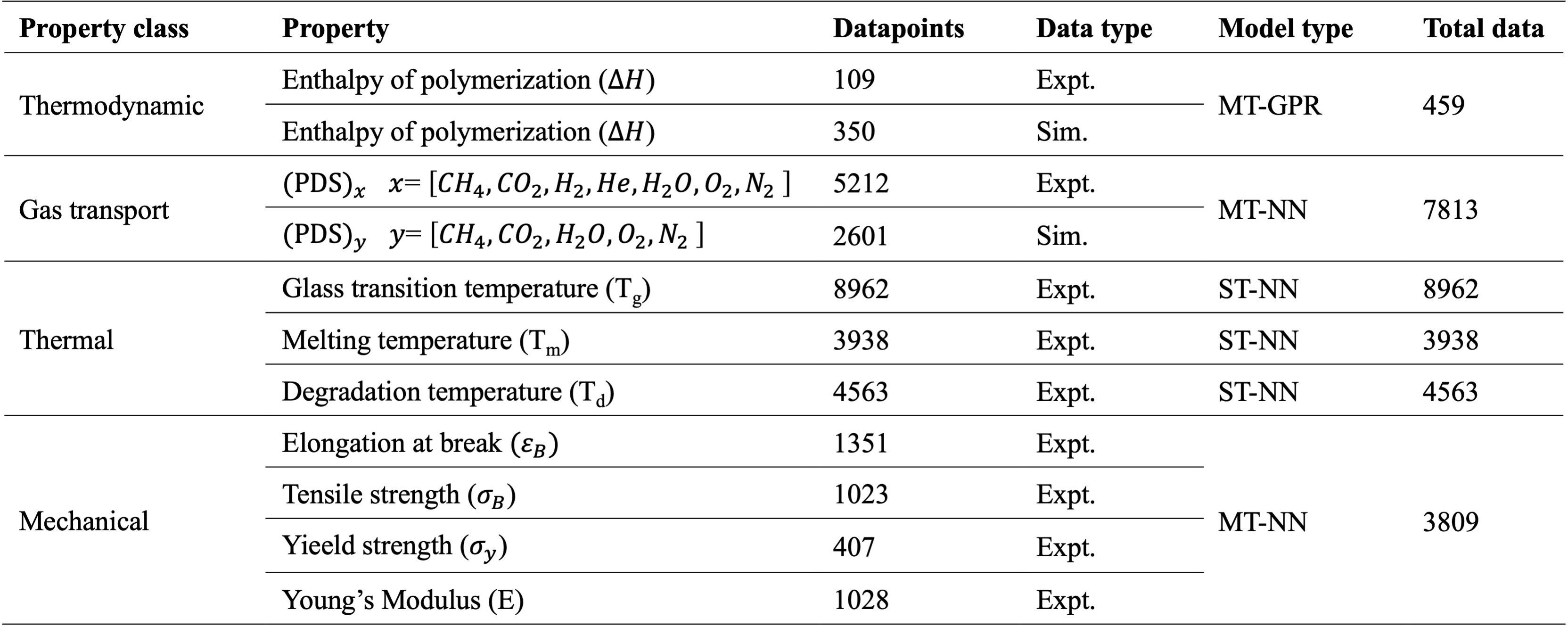}
    \caption{Summary of datasets and model training setup.}\label{fig:datasets}
\end{table}

\begin{table}[h!]
    \centering
    \includegraphics[width=0.8\columnwidth]{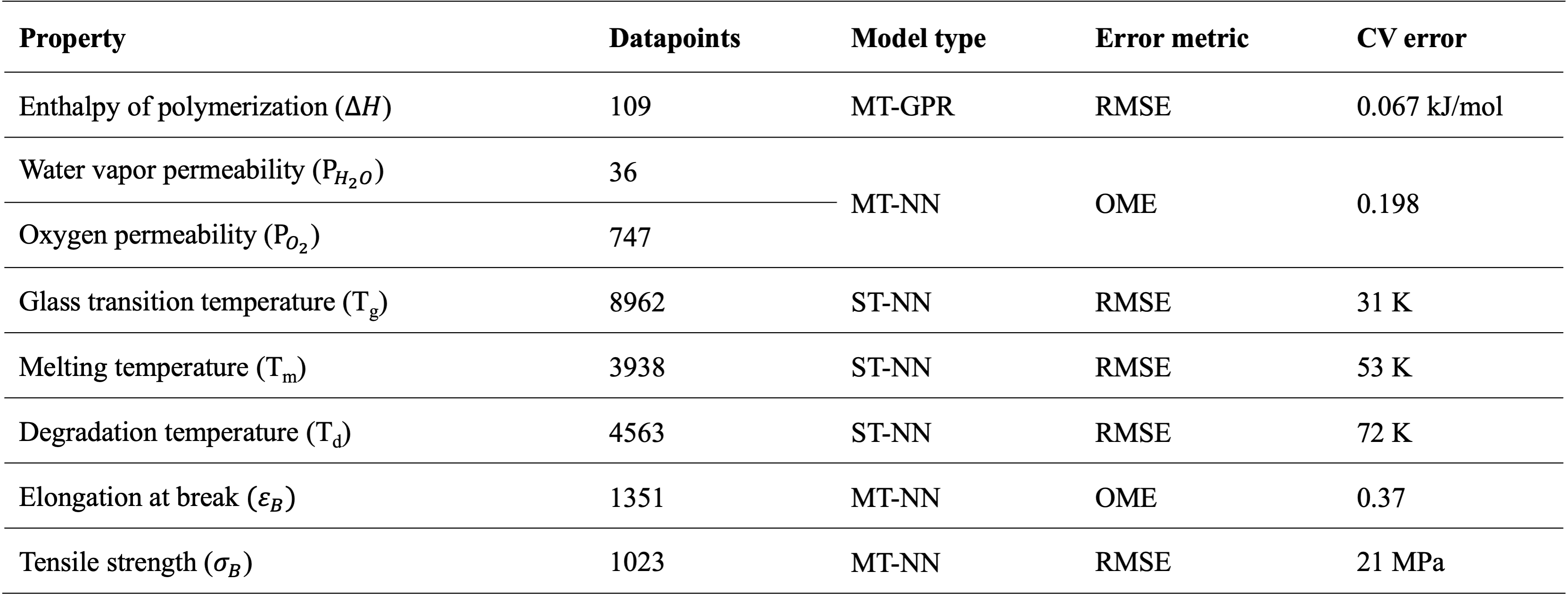}
    \caption{Summary of model performance.}\label{fig:performance}
\end{table}

Our previous work assessed the effects of data fusion on the performative capabilities of permeability of common gases.\cite{phan2024gas} Here, we extend that body of work, following a similar analysis but with a particular focus on the prediction of water permeability. ST and MT models developed with PolymRize were used to predict the water permeability of 39 data points spanning 27 polymers, at various holdout train and test splits. These splits used stratified sampling based on the polymer SMILES string\cite{weininger1988smiles}, giving a sense for how well the model forecasts unseen polymers. In addition, four random seed selections of the splits were used for the computation of the statistics of the model performance. The performance of the models was appraised using the coefficient of determination ($R^2$) and the order of magnitude error (OME).

\begin{figure}[h!]
    \centering
    \includegraphics[width=1\columnwidth]{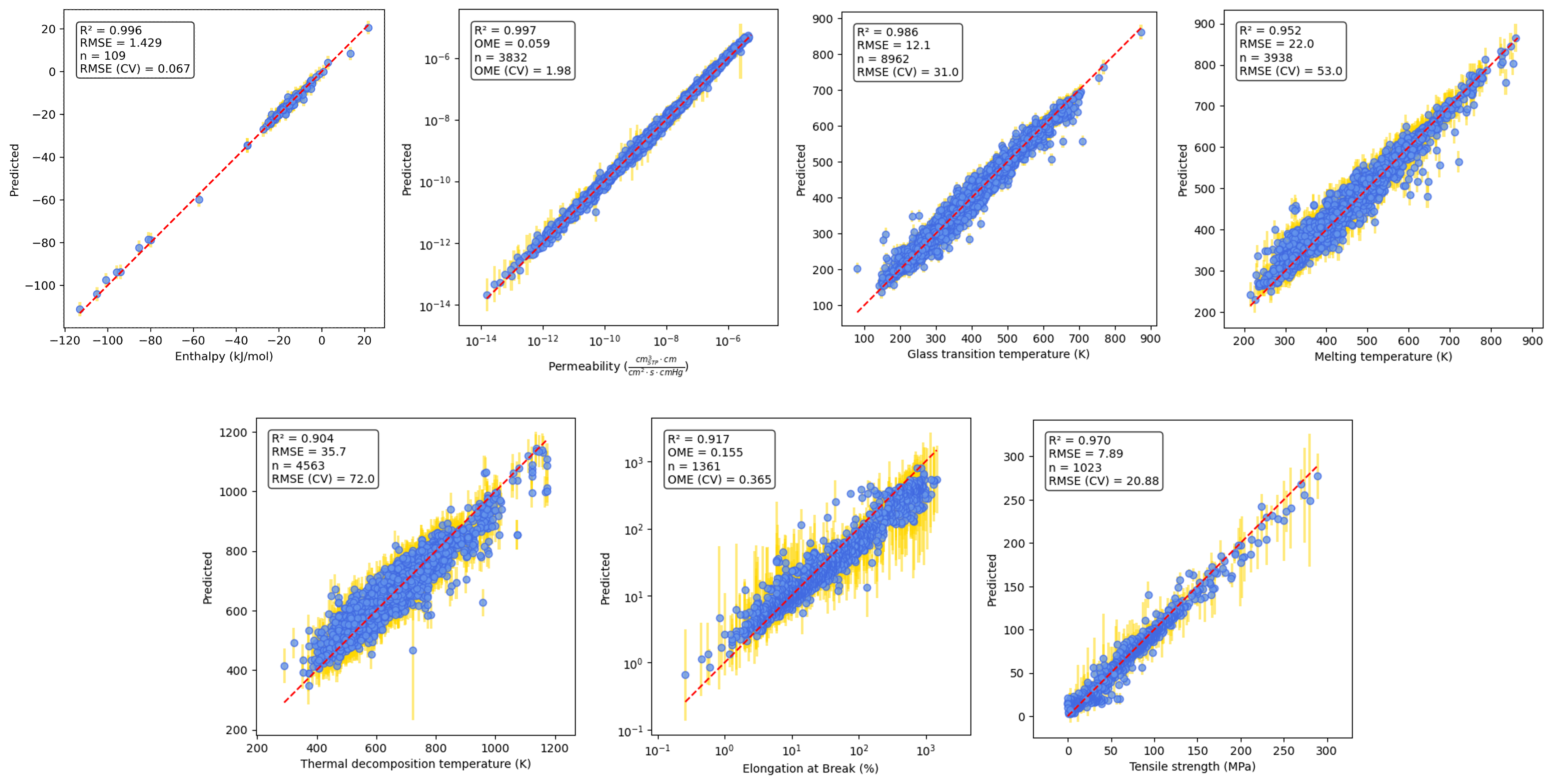}
    \caption{Parity plots of developed ML models. CV: Cross-validation}\label{fig:parity}
\end{figure}

\section{Experimental validation}

\subsection{Synthesis of poly-PDO from PDO}
Ring-opening polymerization of PDO was performed in the bulk with magnetic stirring in oven-dried 20 mL vials. PDO (5.0 g, 49 mmol) was charged into a vial. The vial was sealed with a rubber stopper. The initiator (benzyl alcohol, 10 $\mu$L, 10 $\mu$mol, 2.0 $\times$ 10$^{-2}$ mol $\%$ relative to monomer) and the catalyst (Sn(Oct)$_2$ solution, 10 $\mu$L, 10 $\mu$mol, 2.0 $\times$ 10$^{-2}$ mol $\%$ relative to monomer), were subsequently injected into the vial through the rubber stopper with a syringe. The vials were then transferred to a silicone oil bath and immersed up to their caps. The vials were pulled from the reaction bath at predetermined intervals, and a small portion of each vial was separated and then dissolved in cooled CDCl$_3$ for reaction quenching and $^1$H nuclear magnetic resonance (\textsuperscript{1}H NMR) measurement. The remaining portion of each vial was dissolved in cooled CHCl$_3$. The resulting PPDO was purified by precipitation from the CHCl$_3$ solution with methanol and dried in vacuo. 
The molecular weight of the resulting sample was 26.6 kg/mol.

The chemical structure and composition of the polymers were confirmed by \textsuperscript{1}H NMR spectroscopy in CHCl$_3$ at room temperature. Spectra were acquired on Bruker Avance 400, 500, or 700 MHz instruments and referenced to the residual solvent signal of CHCl\textsubscript{3} at $\delta$ 7.26 ppm. Characteristic resonances for both comonomer units were clearly observed in the \textsuperscript{1}H NMR spectra (Fig. S1). Size-exclusion chromatography (SEC) was performed on a Tosoh EcoSEC HLC-8320 GPC system equipped with TSKgel SuperHZ-L columns using CHCl\textsubscript{3} containing 0.25\% NEt\textsubscript{3} as the eluent at a flow rate of 0.45 mL min\textsuperscript{-1}. Number-average molecular weights (\textit{M}\textsubscript{n}) and dispersities (\textit{Đ}) were determined from refractive-index chromatograms and calibrated against PStQuick Mp-M polystyrene standards.

\begin{figure}[h!]
    \centering
    \includegraphics[width=0.8\columnwidth]{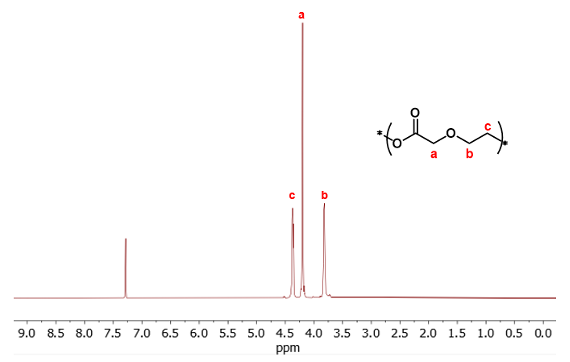}
    \caption{$^1$H NMR spectra of poly-PDO in CDCl$_3$.}\label{fig:NMR1}
\end{figure}

\subsection{Depolymerization of poly-PDO}
An oven-dried 5 mm thick wall NMR sealed sample was charged with poly-PDO (30 mg), DBU (10 mol\% relative to repeating units), and Benzene-d6 (0.5 ml). The NMR tube was sealed by a Teflon stopcock and then immersed in a silicone oil bath at 79 $\degree$C for 6 hours before subjecting to $^1$H NMR measurements. We integrate the methylene of the monomer and the polymer to obtain the integration, from which the degree of degradation was estimated.

\begin{figure}[h!]
    \centering
    \includegraphics[width=0.8\columnwidth]{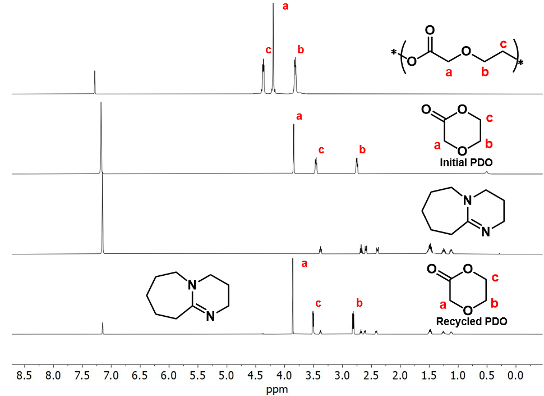}
    \caption{Overlays of $^1$H NMR spectra of initial and recycled PDO and poly-PDO in benzene-d6.}\label{fig:NMR2}
\end{figure}

\subsection{Gas transport measurements}

The poly-PDO membrane film for the gas transport measurements was prepared by first dissolving 1 wt\% of the polymer in chloroform, and then filtered using 0.22 $\mu m$ PTFE syringe filters (VWR), and then cooled to 277 K. A Matrimid support was fabricated following the procedure reported in the literature.\cite{lee2025prediction} 0.5 $mL$ of the dope solution was then spin-coated onto a cross-linked Matrimid support at 1,000 rpm for 2 minutes, resulting in an approximately 300 $nm$ thin film of poly-PDO.

\begin{figure}[h!]
    \centering
    \includegraphics[width=.9\columnwidth]{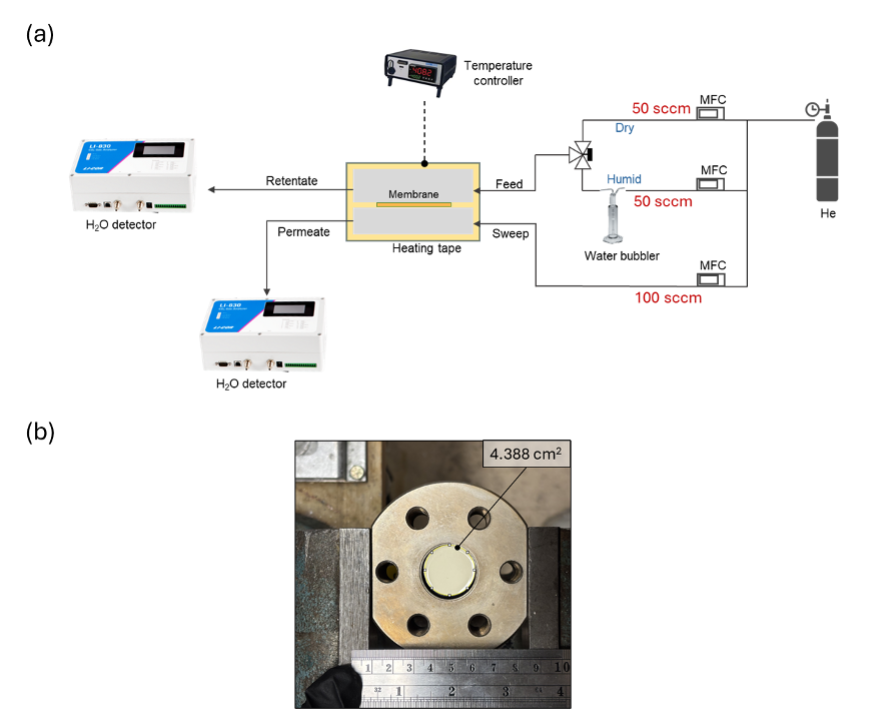}
    \caption{Gas permeability measurement setup. (a) Wicke–Kallenbach cell. (b) Top view of the membrane cell.}\label{fig:setup}
\end{figure}

Oxygen permeation of the poly-PDO membrane was evaluated using a constant-pressure, single-gas permeation setup. A membrane sample with an effective area of 14.5 $cm^2$ was mounted in the sample cell, and both the upstream and downstream sides were evacuated with oxygen for 20 minutes. After evacuation, the upstream side and gas reservoir were isolated by closing the vent valve, and the reservoir was pressurized to 40 $psi$ with pure oxygen at room temperature. The gas inlet valve was then opened, and the oxygen flow rate on the permeate side was determined by tracking the movement of a bubble through a 10 $mL$ graduated pipette filled with Snoop solution. Each measurement was repeated five times, and the average value was used to calculate the oxygen permeability of the poly-PDO.

Water permeation experiments were conducted using a Wicke–Kallenbach cell with a poly-PDO (Fig. \ref{fig:setup}a). Helium (He) was used as the carrier gas and divided into three streams, each regulated by an individual mass flow controller (MFC). Two of the streams were combined to form the feed gas: one was passed through a water bubbler to generate humidified He, while the other supplied dry He. A three-way valve enabled mixing of the two streams, and the feed humidity was adjusted by varying the dry-to-humid flow ratio. The third stream provided dry He as the sweep gas. Both feed and sweep gases entered the membrane cell, where the PDO membrane was mounted. The retentate and permeate streams were directed to LI-850 gas analyzers (LI-COR Biosciences), which continuously measured $CO_{2}$ and $H_{2}O$ reported concentrations. The membrane cell temperature was regulated by a thermocouple connected to a proportional–integral–derivative (PID) controller, ensuring stable operation at the target temperature. A top view of the membrane cell is shown in Fig. \ref{fig:setup}b. The effective membrane area was 4.388 $cm^2$, determined using ImageJ analysis of the mounted membrane image.

\subsection{Thermal measurements}
The glass transition temperature was measured using differential scanning calorimetry (DSC) under a nitrogen atmosphere with a heating rate of 10 $\degree$C/min, on a Mettler Toledo DSC 3+ (STARe).

\begin{figure}[h!]
    \centering
    \includegraphics[width=0.8\columnwidth]{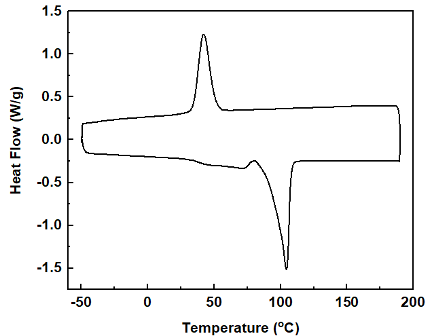}
    \caption{DSC curve of poly-PDO.}\label{fig:DSC}
\end{figure}

\subsection{Mechanical measurements}

The PDO films for mechanical measurements were formed via compression molding using a manual benchtop Carver press and sectioned into rectangular specimens following ASTM D882 guidelines. Before pressing, the material was preheated between the plates at 110 $\degree$C for 10 min to ensure uniform thermal equilibration. Pressure was then applied and maintained for an additional 15 min, facilitating complete melting and consolidation of the polymer. The resulting films possessed a nominal thickness of approximately 1 mm. Following molding, the films were cooled under ambient laboratory conditions in air. Uniaxial tensile tests were conducted on a universal test machine (Insight 10, MTS Systems Corp., Eden Prairie, MN, USA) at a constant crosshead displacement rate of 5 $mm$*$min^-1$.

\newpage
\bibliography{refs}